\documentclass[aps,prb,twocolumn,superscriptaddress,showpacs,ams]{revtex4}

\usepackage{graphicx}
\usepackage{bm}
\usepackage{amssymb} 
\usepackage{psfrag,color} 

\newcommand{\beq}   {\begin{equation}}
\newcommand{\eeq}   {\end{equation}}
\newcommand{\ba}   {\begin{eqnarray}}
\newcommand{\ea}   {\end{eqnarray}}
\newcommand{\maxim}   {\mbox{\scriptsize max}}
\newcommand{\onedot}   {\mbox{\scriptsize Dot}}

\newcommand{\dotleads}   {\mbox{\scriptsize Dot-Leads}}
\newcommand{\leads}   {\mbox{\scriptsize Leads}}

\newcommand{\st}   {\mbox{\scriptsize st}}

\begin{document}

\title{Transport properties and Kondo correlations in nanostructures:
the time-dependent DMRG method applied to quantum dots coupled to Wilson chains}

\author{Luis G.~G.~V. Dias da Silva}
\affiliation{Materials Science and Technology Division,
Oak Ridge National Laboratory, Oak Ridge, Tennessee 37831, USA and\\
Department of Physics and Astronomy, University of Tennessee, Knoxville, Tennessee 37996, USA}
\author{F. Heidrich-Meisner}
\affiliation{Institut f\"ur Theoretische Physik C, RWTH Aachen University, 52056 Aachen, and JARA -- 
Fundamentals of Future Information Technology, Germany}
\author{A.~E. Feiguin}
\affiliation{Microsoft Project Q, University of California, Santa Barbara, CA 93106, USA}
\affiliation{Condensed Matter Theory Center, Department of Physics, University of Maryland, College Park, Maryland 20742, USA }
\author{C.~A. B\"usser}
\affiliation{Department of Physics, Oakland University, Rochester, Michigan 48309, USA}
\author{G.~B. Martins}
\affiliation{Department of Physics, Oakland University, Rochester, Michigan 48309, USA}
\author{E.~V. Anda}
\affiliation{Departamento de  F\'{\i}sica, Pontificia Universidade
Cat\'olica do Rio de Janeiro (PUC-Rio), 22452-970, Caixa Postal:
38071 Rio de Janeiro, Brazil.}

\author{E.~Dagotto}
\affiliation{Materials Science and Technology Division,
Oak Ridge National Laboratory, Oak Ridge, Tennessee 37831, USA and\\
Department of Physics and Astronomy, University of Tennessee, Knoxville, Tennessee 37996, USA}
\date{\today}

\begin{abstract}

We apply  the  adaptive time-dependent Density  Matrix Renormalization
Group method  (tDMRG)    to the study   of   transport  properties  of
quantum-dot systems  connected to metallic leads.  Finite-size effects
make  the usual  tDMRG description of   the Kondo regime a numerically
demanding   task.  We  show that  such   effects can be attenuated  by
describing the leads by ``Wilson chains",  in which the hopping matrix
elements decay exponentially away from the impurity ($t_n \propto
\Lambda^{-n/2}$). For a  given system size and in the linear response 
regime, results  for $\Lambda > 1$  show several improvements over the
undamped, $\Lambda=1$ case: perfect  conductance is obtained deeper in
the strongly interacting  regime  and  current  plateaus remain   well
defined for longer time scales.  Similar improvements were obtained in
the finite-bias regime up  to bias voltages of the  order of the Kondo
temperature.  These results show that, with the proposed modification,
the  tDMRG  characterization of  Kondo  correlations in  the transport
properties can be substantially  improved, while it   turns out to  be
sufficient to  work  with much  smaller  system sizes. We discuss  the
numerical cost of  this approach with respect  to the necessary system
sizes and the entanglement growth during the time-evolution.

\end{abstract}


\maketitle


\section{Introduction}
\label{sec: Intro}

The current excitement in the condensed matter and materials science
community  surrounding the study of nanoscale transport stems from both the
potential applicability in molecular electronic devices \cite{joachim00} and
the possibility of designing nanostructures to realize quantum impurity
Hamiltonians. Hallmark experimental achievements include the observation of
the Kondo effect in quantum dots,\cite{goldhabergordon98,wiel00} molecules,\cite{park02}
nanotubes,\cite{jarilloherrero05,makarovski07}
and non-Fermi liquid behavior in quantum dot structures.\cite{potok07}

While transport is an intrinsic non-equilibrium situation, in the linear
response  regime  transport coefficients  are  commonly derived  from
equilibrium  correlation  functions.\cite{mahan,  hewson}
A prominent numerical  tool for describing the  equilibrium Kondo regime
is  Wilson's  numerical  renormalization-group  method (NRG).  In  the
original NRG  formulation for the Kondo  model,\cite{wilson83,bulla08} Wilson
showed that  the contribution from band states  exponentially close to
the Fermi  energy needs to be  taken into account in  order to capture
the  correct properties of the  ground  state.  For  this  reason,
standard  tight-binding numerical  approaches  face  a  formidable
challenge in  addressing  this problem: finite-size effects set
a minimum energy scale,  the level spacing,
below which the calculation cannot  capture the crossover to the Kondo
state.\cite{thimm99,affleck01,hand06}

Wilson proposed a combination of  two elements to handle this
problem: (i) A discretization procedure of the metallic band,
leading to a mapping into a  impurity connected  to a
one-dimensional  tight-binding chain with  exponentially decaying
hoppings. We  will refer to such  leads as {\it Wilson chains} in
this work.  (ii)   A   non-perturbative renormalization procedure
that probes  successive  energy scales  by recursively
diagonalizing the  Hamiltonian and  keeping  the relevant states
at each scale.

Recent theoretical
\cite{wingreen93,schoeller00,schiller00,rosch01,fujii03,kehrein05,anders05,alhassanieh06,doyon06,schneider06,mehta06,doyon07,jakobs07,anders07,anders08,kirino08,weiss08}
and experimental efforts \cite{grobis07} aim at observing and modeling
genuine non-equilibrium physics.  A particularly important question is
under what conditions   steady-state  situations  can be   reached  in
numerical simulations, and promising results  have been obtained using
time-dependent
approaches.\cite{alhassanieh06,schneider06,anders05,kirino08,anders07,anders08}
Such  ideas   have  been  pursued  using   both   the density   matrix
renormalization          group           (DMRG)              technique
\cite{white92b,white93,schollwoeck05,hallberg06}      and          the
NRG,\cite{bulla08,anders05,anders07,anders08}   in   the  former  case  utilizing   the  adaptive
time-dependent DMRG.\cite{daley04,white04}

Moreover, the incorporation  of ingredients of DMRG  into NRG and vice
versa    has  led     to    a   significant   extension     of    both
methods.\cite{hofstetter00,nishimoto04,anders05,weichselbaum08,saberi08,holzner08}
As a prominent example, this includes the use of Wilson chains in DMRG
for the description of the Kondo regime of the Anderson impurity model
in  Refs.~\onlinecite{nishimoto04,weichselbaum08},     where        in
Ref.~\onlinecite{weichselbaum08} the  common mathematical structure of
NRG and (single-site) DMRG in terms of  matrix-product states has been
exploited. Recently, a  similar idea  has been  successfully  explored
within a cluster-embedding approach,  resulting in the development  of
the   so-called    logarithmic   discretization     embedded   cluster
approximation (LDECA).\cite{anda08}

The advantage of DMRG is  its flexibility: it is in principle
possible to  model  complex  interacting  regions
\cite{schollwoeck05}  or  to incorporate interactions into the
leads (see, {\it e.g.}, Ref.~\onlinecite{costamagna06}).  Moreover, it
is the numerical  method of choice  for  one-dimensional bulk systems,
and it allows for the calculation of extended correlation functions in
a straightforward way.   For the description of  transport  phenomena,
there is no  restriction to work in the  small bias  regime, as finite
biases         can     be       incorporated     into   time-dependent
simulations.\cite{alhassanieh06,schneider06,kirino08} 

In  transport  investigations  based  on  DMRG,   several groups  have
introduced modifications in the contact leads, such as the logarithmic
discretization,    to improve the      results  of either  the  ground
state\cite{nishimoto04,weichselbaum08} or  tDMRG  calculations.  These
also include damped boundary conditions\cite{bohr06,alhassanieh06} and
a momentum-space  representation   of the leads.\cite{bohr07} In   the
latter work, an interacting resonant level model has been studied with
tDMRG and, effectively, a logarithmic discretization  of the leads has
been used.  Working with different kind of leads, while preserving the
main physical  properties  of the system,  has  thus  proven to   be a
promising path that we will further pursue in this work.

It is the  purpose  of this paper   to perform  a numerical  real-time
analysis of Kondo correlations in quantum dot problems using tDMRG and
Wilson chains.  We show that, compared to a  previous study by some of
us,\cite{alhassanieh06}  a correct description  of transport through a
quantum dot  can be obtained deeper  into the Kondo  regime, and using
much smaller system sizes. While a tDMRG analysis  of the Kondo regime
based on a real-space  description is hampered by finite-size  effects
in the leads, we show that an appropriate choice of hopping amplitudes
in the leads nicely  circumvents such problems.

In this  sense, the  discretization scheme  proposed  by  Wilson is  a
natural  choice:  the  noninteracting tight-binding   chain becomes an
effectively metallic system with reduced level spacings, corresponding
to a  subset   of states  directly  coupled  to  the impurity   in the
continuum limit.  \cite{hewson,wilson83} In  addition, it turns out to
offer advantages in  the   time-dependent description as well:   for a
given system size, it substantially  increases the characteristic time
scales  over which a constant current  flows between the leads, before
reaching the system's boundary.  This  turns out to be crucial  deeper
into   the  Kondo  regime  (i.e., for  small   Kondo temperatures), as
constant currents sustained over  longer time scales are necessary  to
access  the regime of coherent  transport  through the sharp, resonant
Kondo state. \cite{wingreen93,nordlander99,schiller00} We also discuss
and study the influence of the discretization  at finite-bias.  In the
low-bias  regime, the logarithmic   discretization gives  the  correct
description  of the Kondo regime,  at   large bias,  (where the  Kondo
effect is effectively destroyed), a linear discretization of the leads
and larger chains should be used.

The paper  is    organized as  follows:   In Sec.~\ref{sec:model}   we
introduce our   approach, using the  example  of a  single quantum-dot
connected to metallic leads out of equilibrium, treated with the tDMRG
method. Results for the  time-dependent  currents and charge  transfer
are discussed in Secs.\ \ref{sec:curr} and
\ref{sec:charge}, respectively.
Particular emphasis is devoted to describing the improvements obtained
by choosing a $\Lambda>1$ model over the  $\Lambda=1$ case, and to the
dependence of our results on both the discretization parameter and the
system size.  Moreover, we discuss the conductance results, taken from
the  time-dependent data, in  Sec.\  \ref{sec:Conductance}.   In Sec.\
\ref{sec:bias}, we   provide  a discussion  of  how  to  approach  the
finite-bias  regime by combining  tDMRG data obtained from discretized
and   tight-binding      leads.     We    present    a   summary    in
Sec.~\ref{sec:sum}.      Two       appendices conclude    this   work:
App.~\ref{app:u0} contains  our  results for  the noninteracting case,
and  App.~\ref{app:comp}  provides a  discussion  on the computational
aspects  of our approach     and the entanglement  growth  during  the
time-evolution.

\section{Model and set-up}
\label{sec:model}

As a case study of the  proposed modification of the method, we apply
tDMRG  to a  model  representing  a single  quantum  dot connected  to
metallic  leads.  The  equilibrium and  linear-response  properties of
this system  are well  known and provide  a natural  benchmark against
which we can compare the tDMRG  results.

\begin{figure}[tbp]
\includegraphics*[height=0.15\columnwidth,width=1.0\columnwidth]{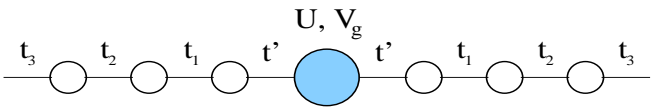}
\caption{(color online) Schematic representation of the model
describing a quantum dot (large circle, blue in the online version) connected to left
and right leads. }
\label{fig:ChainLambda}
\end{figure}

The  quantum  dot  is  modeled  by  a Hubbard  site  with  an  on-site
interaction $U$  and a gate potential $V_g$  coupled to noninteracting
tight-binding chains, representing the leads, as depicted in Fig.
\ref{fig:ChainLambda}.   The  dot-lead   couplings   are  labeled   by
$t^{\prime}$.
The full Hamiltonian reads
$H=H_{\onedot}+H_{\dotleads}+H_{\leads}$, with
\begin{eqnarray}
H_{\onedot} & = &V_g \hat{n}_{0} + U\hat{n}_{0
\downarrow}\hat{n}_{0 \uparrow} \nonumber \; ,\\
H_{\dotleads} & = & -t^{\prime} \sum_{\sigma, \alpha=R,L} \left( c^{\dagger}_{0,\sigma} c_{1, \alpha, \sigma} + \mbox{h.c.} \right)\nonumber \; ,\\
H_{\leads} & = & -\sum_{\sigma, \alpha=R,L}\sum_{n=1}^{N_{\alpha}}
t_n \left( c^{\dagger}_{n, \alpha, \sigma} c_{n+1, \alpha, \sigma} +
\mbox{h.c.} \right) , \nonumber \\
\label{Eq:Hami}
\end{eqnarray}
where $c^{\dagger}_{n, \alpha, \sigma}$ creates an electron with
spin $\sigma$ at site $n$ in lead $\alpha$ ($n=0$ is the dot site),
 $N_{\alpha}$ counts the number of sites in lead $\alpha$,
and $\hat{n}_{i\sigma}=c^{\dagger}_{i,\sigma} c^{}_{i,\sigma}$; $\hat{n}_{i}=\hat{n}_{i\uparrow}+\hat{n}_{i\downarrow}$
with $i=0,\{n,\alpha\}$.
The total number of sites is thus $N=N_{L}+1+N_{R}$. Our
results were obtained using
``even-1-odd'' chains, with $N_{L}=N/2$, $N_{R}=N/2-1$.
 Note that finite-size effects due to different cluster types, as discussed in Ref.~\onlinecite{hm08},
vanish at sufficiently large $\Lambda$.

In the spirit of Wilson's discretization scheme, we consider the
hopping matrix elements $t_n$ in the leads to decay exponentially as
\begin{equation}
t_n=t_0\,\Lambda^{-n/2} \,,
\end{equation} where $\Lambda\geq 1$ is the ``discretization
parameter" from Wilson's original formulation.

Unless  otherwise  noted, in  the  following  we  set $t_0=1$,  $U=1$,
$V_g=-U/2$ (particle-hole symmetric point). We focus on results  for $N=32$ sites for
$\Lambda>1$ and up to $N=128$  sites for $\Lambda=1$.  We usually work
at  half-filling of  the whole  system.   In the  range of  parameters
considered, the equilibrium ground-state properties for $\Lambda>1$ do
not  change  significantly  upon   increasing  $N$  beyond  a  certain
$N^*(\Lambda)<50$ at a given $\Lambda$, as we have numerically verified
for a  few cases.  This  is consistent with  NRG runs for  an Anderson
model with similar parameters, which  reach the Kondo fixed point with
less than 50 iterations for $\Lambda\sim 2-3$.

Details    on     the    tDMRG     can    be    found     in    Refs.\
\onlinecite{white04,daley04}.  We  use a Trotter--Suzuki breakup
of  the time-evolution  operator  and typical  time  steps of  $\delta
\tau=0.01$--$0.1$.  A larger time-step $\delta \tau\sim 0.4$ is
sufficient when deeper in the Kondo  regime (large $U/{t^{\prime}}^2$) as resonant
transport is dominated by a small energy scale, the Kondo temperature,
corresponding to long time  scales. \cite{nordlander99,schiller00}
Note  that we denote  time by  the symbol $\tau$ and it is measured in units of
$\hbar/t_0$.
The truncated  weight $\delta \rho$ during  the time-evolution  is typically  kept below
$10^{-7}$ (see App.~\ref{app:comp} for a discussion).

In order to drive a current,  we first compute the ground state of the
system  without a  bias. The  bias is  applied  as
\begin{equation}
H_{\mathrm{bias}}= \frac{\Delta V}{2}\sum_{n=1}^{N_R}
\hat{n}_{R,n} -      \frac{\Delta V}{2}\sum_{n=1}^{N_L}
\hat{n}_{L,n}
\end{equation}
 in the leads  at  time $\tau$ and we  then  evolve in time  under the
 dynamics of $H+H_{\mathrm{bias}}$. We typically work  at a small bias
 of  $\Delta   V=0.005$   (the  finite-bias   case   is  discussed  in
 Sec. \ref{sec:bias}).   The  current  $J(\tau)$ is measured   as  the
 average over the expectation values of the local current operator
\begin{equation}
\hat{J}_{1,\alpha} = i t'\sum_{\sigma} (c_{0,\sigma}^{\dagger}
c_{1,\alpha,\sigma}-\mbox{h.c.}),
\label{eq:j}
\end{equation}
on the links connecting the dot to the leads. This means that we first take
its  expectation value  in the  time-dependent wave-function  and then
average over the two local currents on the links directly connected to
the dot.  We have tried other spatial forms for the applied
bias, {\it e.g.},  a broadened step function.\cite{white04} This  mostly affects the
short-time transient  behavior, but leaves unaffected the average value  of the
current taken  over time (see  Sec.~\ref{sec:Conductance}).
Further    details    on    the     set-up    can    be    found    in
Ref.~\onlinecite{alhassanieh06}.

\section{Time-dependent currents}
\label{sec:curr}
\begin{figure}[tbp]
\includegraphics[width=1.0\columnwidth]{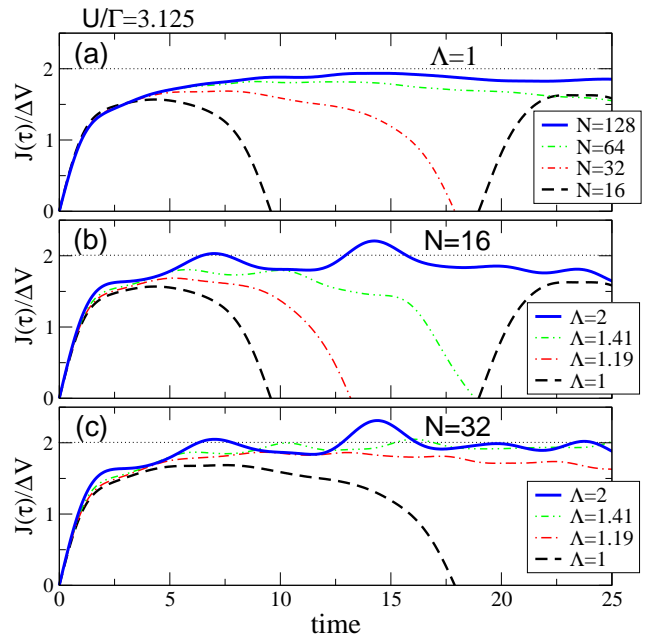}
\caption{(color online) Conductance $J(\tau)/\Delta V$ for $U=1$, $t^{\prime}=0.4$,
$V_g=-U/2$. (a) Fixed $\Lambda(=1)$, and $N=16$, $32$, $64$ and $128$; (b),(c):
Fixed $N$. (b) $N=16$ and (c) $N=32$ with $\Lambda=1$, $2^{1/4}$, $\sqrt{2}$, and $2$.
}
\label{fig:G_NLambda}
\end{figure}

Figure \ref{fig:G_NLambda} shows the current $J(\tau)$ (in units of $e^2/h$) as a
function of time  and divided by the external bias $\Delta V$, at
$U=1$, $t'=0.4$. This corresponds to a ratio of $U/\Gamma=3.125$,
where
$\Gamma=\pi\rho_{\mathrm{leads}}(t^{\prime})^2$ is the
hybridization parameter and $\rho_{\mathrm{leads}}=2/(\pi t_0)$ is
the density of states of the leads in the limit of $\Lambda=1$ and long
chains.

Figure~\ref{fig:G_NLambda}(a) contains the results  for
$N=16,32,64,128$ at $\Lambda=1$, reproducing those of
Ref.~\onlinecite{alhassanieh06}. The other two panels display
$J(\tau)/\Delta V$ for (b) $N=16$ and (c) $N=32$, computed with
$\Lambda=1,2^{1/4},\sqrt{2}$, and $2$.

The   comparison  between   Fig.~\ref{fig:G_NLambda}(a)  and   (c)
is revealing.  In  the $\Lambda=1$ case, the  conductance plateaus
become longer and  higher as the system  size increases
\cite{alhassanieh06}
with the ``plateau length" ($\equiv \tau_{\st}$) increasing
linearly with $N$.
For  $N=128$   sites,  a  nearly  perfect conductance plateau with
$G(\tau)\approx G_0$  is obtained for these parameters. A
finite-size scaling analysis, done in
Ref.~\onlinecite{alhassanieh06} for $U/\Gamma=3.125$, shows that
$G \rightarrow G_0$ for $N^{-1} \rightarrow    0$.

As previously argued  in Ref.~\onlinecite{alhassanieh06}, this plateau
signals the  formation of a  Kondo state in  the system formed  by the
quantum dot and the leads.  A similar Kondo conductance plateau can be
obtained by increasing $\Lambda$ while keeping the system size
\textit{constant}, as shown in Figs.\ \ref{fig:G_NLambda}(b) and (c).
Interestingly, for $N=32$, a plateau with an average conductance of $G
\approx G_0$ can be obtained with $\Lambda=2$.  Thus, an accurate
description of the Kondo regime can be obtained using a relatively
small system by taking $\Lambda>1$.

Notice that  the  $J(\tau)/\Delta V$ curve for $\Lambda=2$ and
$N=32$ is similar to the corresponding
one for $N=16$: an increase in system size had little effect in
the current in this case, as opposed to the $\Lambda=1$ curves.
This weak dependence of $J(\tau)$ on $N$ for large enough $N$ and
$\Lambda$ is a general feature of the $\Lambda>1$
case, and it is a consequence of the exponential decrease of the
hopping matrix elements $t_n$ at large $n$.

This allows us to formulate some key results that can be inferred
from this plot: (i) By comparison of Fig.~\ref{fig:G_NLambda}~(a)
and Fig.~\ref{fig:G_NLambda}~(c), we find that with discretized
leads, a four times smaller system size is sufficient to obtain
roughly the same average current. (ii) At $\Lambda>1$ and for a
given $N$, the current plateaus are longer in time and, on the
time-scales simulated, we do not observe a recurrence (bouncing
current) in the case of $N=32$. (iii) Oscillations about the
average value of the current tend to increase with $\Lambda$.

The small time-scale oscillations are a general feature of the
$\Lambda>1$ case, which emerges in the $U=0$ case as well, as
we have verified with exact diagonalization (see App.~\ref{app:u0}).
We have conducted extensive checks to rule out either
the truncation error during the time evolution or the size of the
time-step  as possible sources of such oscillations at
finite $U$. In fact, the position of the peaks and valleys of the
oscillations as seen in Figs.~\ref{fig:G_NLambda}(b) and (c)
 are $\Lambda$-dependent:
for a given $\Lambda$, the position of
these peaks and valleys remains practically constant even when
other parameters, such as $t^{\prime}$ or $V_g$, are modified,
while it changes  as $\Lambda$ is varied (see
App.~\ref{app:u0}).

The oscillations are reminiscent  of the so called ``current
ringing'' in  mesoscopic transport.\cite{wingreen93}  Similar
current  ringing effects   have  been  observed   in  previous
tDMRG   studies  of noninteracting   systems   away  from
half-filling   or  at   finite bias.
\cite{white04,cazalilla02,schneider06}  In  the   present  case of
both a finite $U$ and $\Lambda>1$, the  oscillations are  present
even at  half-filling and become more prominent at larger values
of $\Lambda$. We stress that the oscillations seen in the
$\Lambda>1$ case are an artefact of the discretization and are
thus not a property of the $\Lambda\to 1$, $N\to \infty$ limit. In
practice, the oscillations can be reduced by using the so-called
z-trick,\cite{bulla08,anders06,oliveira94} which  we illustrate in
App.~\ref{app:u0} for the noninteracting case.

We attribute the increase of the average current on small chains
at larger $\Lambda$ to a combination of mainly two factors: (i) an
effective reduction in the mean-level spacing in the leads in
equilibrium and, quite importantly, (ii) an increase in the
duration of the non-equilibrium current plateaus
above a characteristic Kondo time scale. As the exponential decay
of hoppings leads to an exponential decrease in the velocity at
which charges move far away from the dot, those charges get trapped
at the leads and a recurrence, {\it i.e.}, a reversal of the
current's sign, is not observed. We will elaborate on this point
in the following section.

Before turning to the calculation of conductances from our time-dependent data,
we will discuss the charge profiles and charge transfer during the time-evolution.

\section{Charge profile and charge transfer}
\label{sec:charge}

\begin{figure}[!t]
\centerline{\includegraphics[width=0.6\columnwidth, angle=-90]{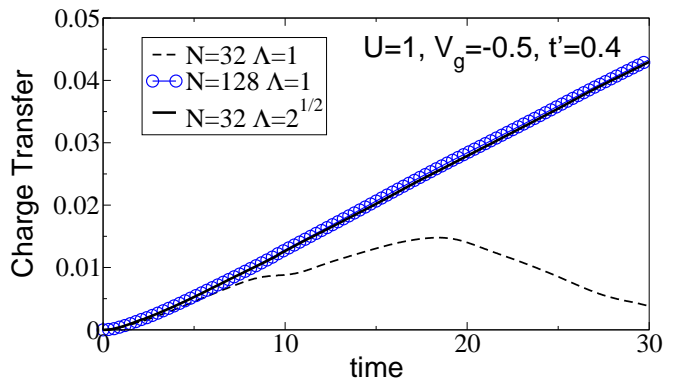}}
\caption{(color online) Charge transfer
for $\Lambda=1$, $N=32$ (dashed lines), $\Lambda=1$, $N=128$ (open
circles) and $\Lambda=\sqrt{2}$, $N=32$ (solid line). The
parameters are $U=1$, $V_g=-U/2$, and $t^{\prime}=0.4$
($U/\Gamma=3.125$). The  maximum in the charge transfer for $N=32,\Lambda=1$ indicates a reversal
in the current.
}
\label{fig:Current_CTrf_1dot}
\end{figure}

Since no dissipative terms are included in the Hamiltonian
(\ref{Eq:Hami}), the total charge is conserved at all times. Thus,
the existence of a net current signals the transfer of charge from
one lead to the other. As time progresses, a saturation point
might be reached, opening the possibility for the current to
decrease and reverse sign, and to transfer  the excess charge
back to the original lead. \cite{alhassanieh06,schneider06}

This mechanism is shown,
for instance, in Fig.\ \ref{fig:G_NLambda}(a) ($\Lambda=1$): for
$N=16$ the current reverses sign around $\tau=9$ while for $N=32$
the sign reversal occurs at $\tau=18$. Such sign reversal also
occurs for $N=128$ and $\Lambda=1$, at times $\tau\gg 25$.

Figure\ \ref{fig:Current_CTrf_1dot} shows the {\it charge transfer} into the left lead
($l$), defined as ($n_i(\tau)=\langle \hat{n}_i(\tau)\rangle$):
\beq
\Delta n_{\mathrm{L}}(\tau) = \sum_{i \in {\mathrm L}} \lbrack n_i(\tau)-n_i(0) \rbrack\; .
\eeq

Notice that this quantity is related to the time-integrated current
through the dot: it hence reaches a maximum whenever the current
changes sign. For $\Lambda=1$ and $N=32$, $\Delta n_l(\tau)$ reaches a
maximum ($\Delta  n_{l,\maxim} \approx  0.01$) at $\tau=18$  while for
$\Lambda=\sqrt{2}$, it increases to about four times that value.
Remarkably, in  order to obtain a similar  charge transfer enhancement
with  ``regular'' leads  ($\Lambda=1$),  a four-fold  increase in  the
system     size     is    needed     (open     circles    in     Fig.\
\ref{fig:Current_CTrf_1dot}).

The approximately linear  increase in $\Delta n_l(\tau)$ is correlated with
the plateau   in  the   current  in   both  cases   (compare   with  Fig.\
\ref{fig:G_NLambda}).  We  identify this as  a general feature  of the
introduction  of the  decaying  hopping matrix  elements  for a  given
system size $N$: an enhanced charge transfer over longer time scales.

\begin{figure}[tbp]
\includegraphics*[clip,angle=-90,width=1.0\columnwidth]{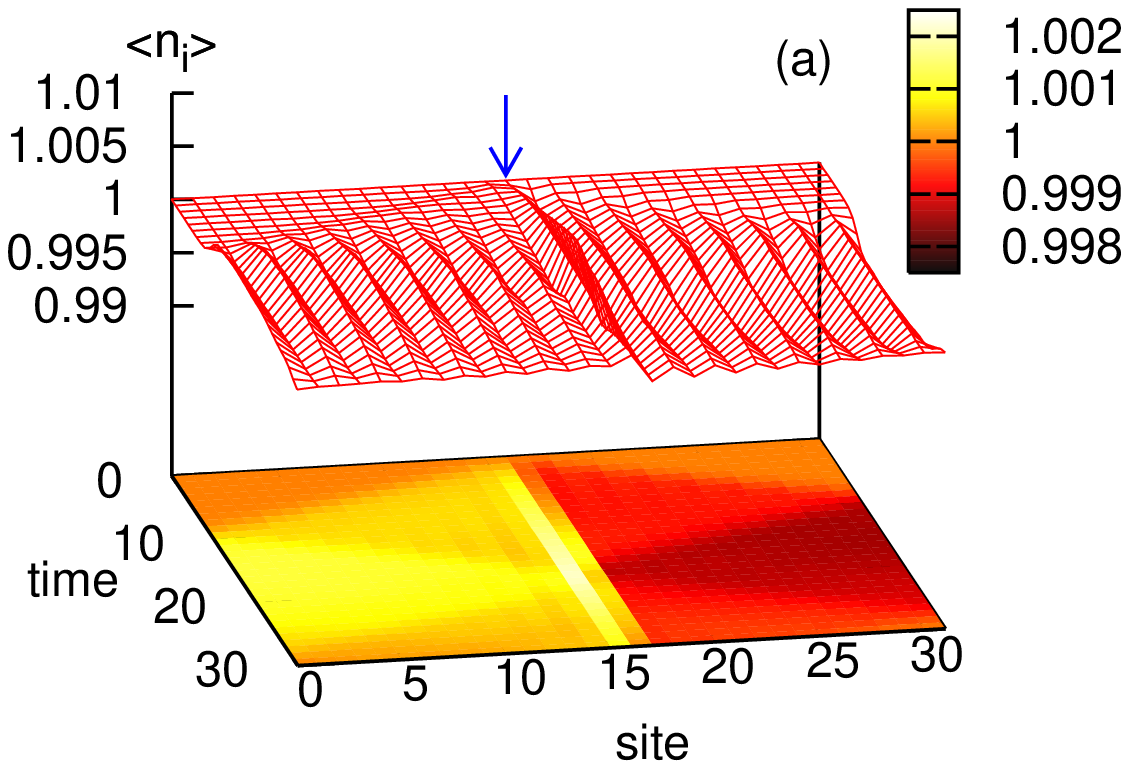}
\includegraphics*[clip,angle=-90,width=1.0\columnwidth]{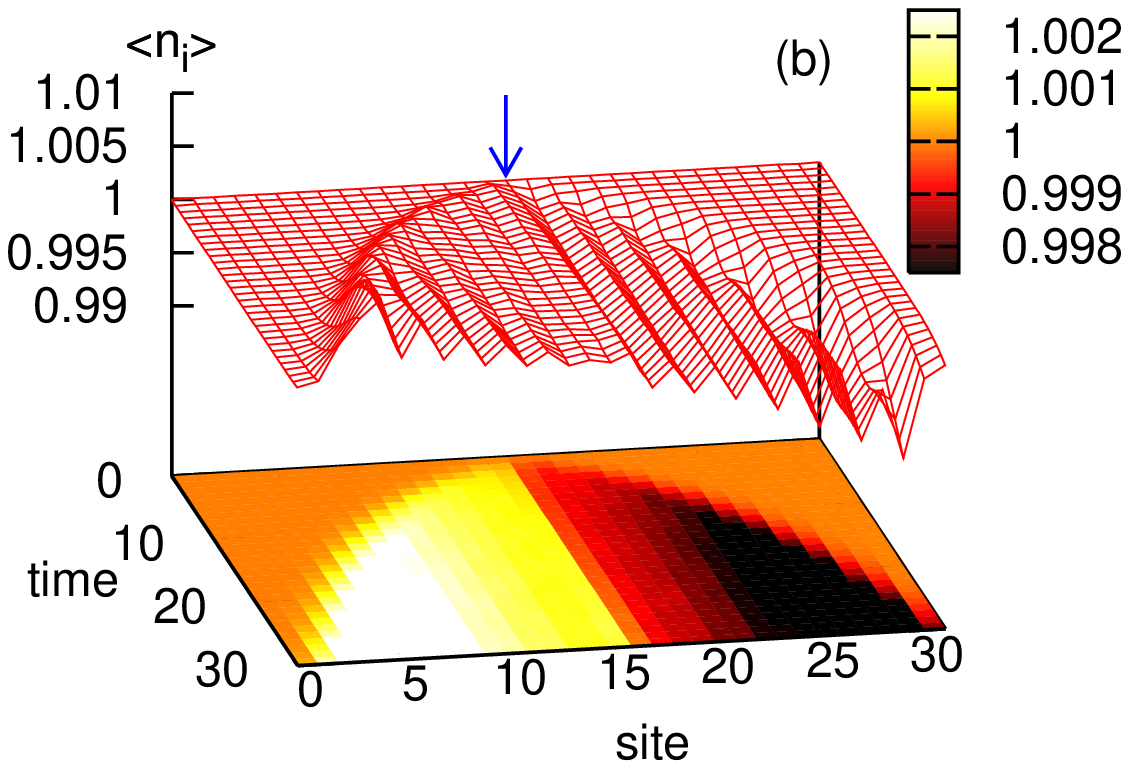}
\caption{ (color online) Charge profile of the 16-1-15 chain
(single dot),
$V_g=-U/2$: (a) $\Lambda=1$ and (b) $\Lambda=\sqrt{2}$. The dot
site is indicated by arrows in both plots. }
\label{fig:ChargeProfile_1dot}
\end{figure}

This indicates  that the exponential  decay in the  hoppings increases
the maximum charge  that can be ``stored'' in the  leads, or, in other
words,  it provides an   increase  in   their  effective  ¸``capacitance''.   As  a
consequence, even  small systems  can hold a  larger amount  of charge
without reversing the current,  leading to longer constant-current
plateaus.

The effect of  $\Lambda>1$  can be
illustrated  by  considering the time  evolution  of  the  charge profile.  Figure
\ref{fig:ChargeProfile_1dot}     shows     the     charge     $n_l(\tau)$
on each  site $l$ of the chain  plotted against time
for a chain of $N=32$ sites. The top panel shows the $\Lambda=1$ case:
charge is  initially transferred from the  left to the  right lead and
back, leading  to an oscillation in $n_l(\tau)$ with a
maximum around  $\tau=18$, as expected.
The charge versus time diagram clearly shows a light-cone with a
wave-front that propagates at the Fermi velocity.
Notice that  the excess charge is mostly accumulated  in the
vicinity of the dot,  which we expect to modify the leads' density
of states
``seen" by the dot.

This is in sharp contrast with the $\Lambda=\sqrt{2}$ case (bottom
panel): a larger charge is transferred to the left lead and no
reflux is noticed. More importantly, the charge tends to
accumulate toward the edge of the leads, with strong Friedel oscillations.
This can be intuitively understood from the
exponential decrease in the couplings: the weak coupling of the
end sites with the remaining of the chain makes them good ``charge
traps''.

\section{The conductance}
\label{sec:Conductance}

We now  turn to the linear conductance $G(\tau) \equiv J(\tau)/\Delta  V$,
expressed in  units of $G_0\equiv 2e^2/h$. We will discuss the dependence
of $G(\tau)$ on  $\Lambda$ and  $N$,  at the  particle-hole symmetric  point.
For this purpose, we refer the reader back to Fig.~\ref{fig:G_NLambda}.
Taking the example of $\Lambda=2$, we see that no significant
difference exists between the conductance curves for $N=16$ and
$N=32$. In general, for a given $\Lambda$ there exists a certain
system size above which no significant changes in
properties, such as the ground state energy or the conductance,
take place as extra sites are added to
 the system.

\begin{figure}[tbp]
\includegraphics[clip,width=1.0\columnwidth]{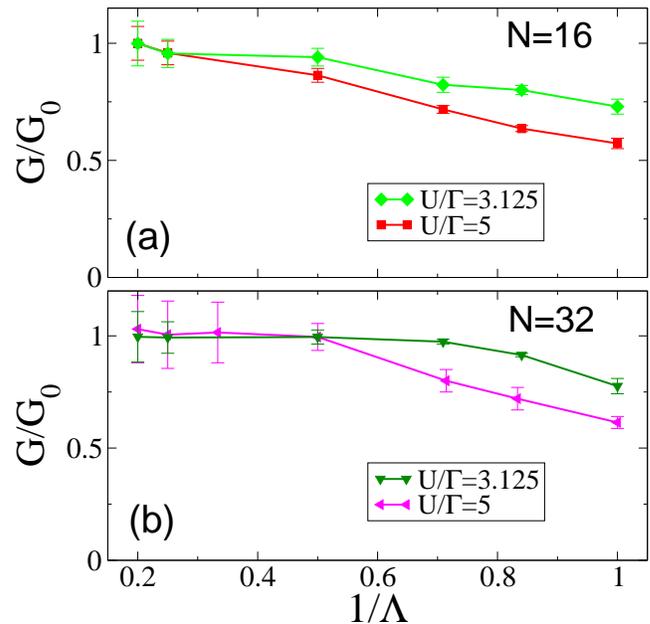}
\vspace{-0.5cm} \caption{ (color online) Scaling with $1/\Lambda$
of the conductance $G$ for (a) $N=16$ and (b) $N=32$,
$V_g=-U/2$,
and different values of $U/\Gamma$.}
\label{fig:Egs_G_Lambda}
\end{figure}

In order to study the  dependence of the conductance with
other  parameters,  we   calculate  the  \textit{average  conductance}
$G=\langle  G(\tau)\rangle_{\tau}$  over a  plateau,  {\it e.g.},  those
displayed  in Fig.\  \ref{fig:G_NLambda}.  This  procedure  carries an
intrinsic  uncertainty which depends  on the truncated weight  in the
DMRG  time  evolution and,  more  importantly,  on  the dispersion  of
$G(\tau)$ around  the average  due to the  current ringing  effects at
larger $\Lambda$.  While  the former can be constrained  below a target value
by increasing  the number  of states  that are kept during the time  evolution (see the
discussion in  App.~\ref{app:comp}),  the latter is intrinsic for $\Lambda>1$.
We estimate such uncertainty by computing
\begin{equation}
\delta G =
\sqrt{\langle                           G(\tau)^2\rangle_{\tau}-\langle
G(\tau)\rangle_{\tau}^2},
\label{eq:deltaG}
\end{equation}
  and  indicate  it  as  error  bars  in  the
figures. We remark  that the main contribution to
$\delta G$ in the plots comes from the current-ringing oscillations (see the analysis in App.~\ref{app:u0}).

Figures~\ref{fig:Egs_G_Lambda}(a) and (b)  show  the scaling  of
$G \equiv \langle  G(\tau)\rangle_{\tau}$ with $1/\Lambda$  for
different values of $U/\Gamma$, and  for $N=16$ and  $32$,
respectively.   The scaling is more conclusive for $N=32$: $G
\rightarrow G_0$ as $1/\Lambda$ decreases, for $U/\Gamma \lesssim
5$. Most importantly, Fig.\ \ref{fig:Egs_G_Lambda}(b) establishes
the convergence of the conductance  in $1/\Lambda$ (obtained at a
fixed system size) to the correct result,  namely perfect
conductance.

\begin{figure}[tbp]
\includegraphics[clip,width=1\columnwidth]{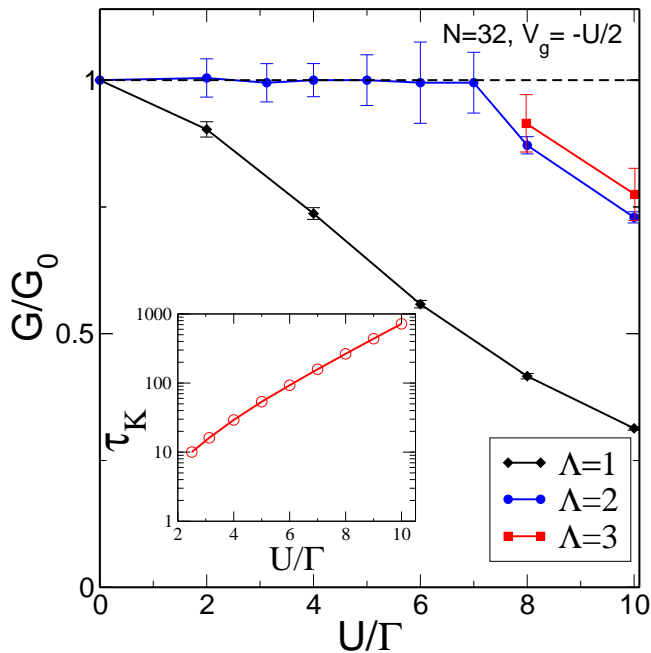}
\caption{ (color online)  Average conductance $G\equiv\langle
G\rangle_{\tau}$ vs. $U/\Gamma$ for $\Lambda=1,2$ and $3$, and
N=32 (dashed line is a guide to the eye). For $\Lambda>1$, the
averages are taken over time intervals $ \tau_K\lesssim \tau
\lesssim \tau_{\mathrm{max}}$. For $U/\Gamma \leq 7$, maximum
simulation times are $\tau_{\mathrm{max}} \approx \tau_K+100$. For
$U/\Gamma\gtrsim 8$ (included for illustration), $\tau_K >
\tau_{\mathrm{max}}$, hence $G<G_0$. Inset: Kondo time $\tau_K$
(calculated with NRG) vs. $U/\Gamma$. See the text for details. }
\label{fig:G_tp_1dot}
\end{figure}

 Figure \ref{fig:G_tp_1dot} depicts results for $G/G_0$
as  a  function  of  $U/\Gamma$,  for $N=32$,  $\Delta  V=0.005$,
and
$\Lambda=1$, $2$,  and $3$. With $\Lambda>1$,  we obtain perfect
conductance  up   to $U/\Gamma  \approx  7$.    This constitutes a
considerable improvement over the $\Lambda=1$ case with $N=32$
(also shown in Fig.~\ref{fig:G_tp_1dot}), for which perfect
conductance plateaus were not observed for nonzero $U/\Gamma$.
Furthermore, we stress that the $\Lambda>1$ approach also gives
more well defined plateaus of constant currents, which in practice
makes easier the averaging of $J(\tau)/\Delta V$ over time.

The improvement previously discussed is anchored on a combination
of two key elements in the $\Lambda>1$ case: (i) an effective
reduction in the level spacing of the  metallic leads in
equilibrium\cite{anda08} and  (ii) the suppression of the current
reversal, which is a consequence  of the reduced velocity in the
leads when $\Lambda>1$. Both points are related to the exponential
decrease in the chain hoppings, which, in Wilson's scheme, can be
traced back to a representation of the continuum of states
directly connected to the quantum dot.

Point  (ii)  is  important  for  the  following  reason:  in  resonant
transport, the  typical time  scale for reaching  the steady  state is
inversely    proportional   to   the    width   of    the   resonance.
\cite{wingreen93,schneider06} The  typical time scale  associated with
the development  of the  perfect conductance plateaus  associated with
the    Kondo   state    is   thus    $\tau_K \equiv \hbar/T_K$.
\cite{nordlander99,schiller00}  It is  then crucial  that  the current
does  not  reverse sign  before  times  of  order $\tau_K$  have  been
reached.

As explained in Sec.\ \ref{sec:curr}, for $\Lambda=1$ the plateaus
last over time intervals
$\tau_{\st} \propto N$ and a compromise must be obtained between
$T_K$ and $N$ such that the condition $\tau_{\st} \gtrsim \tau_K$
is fulfilled. For $\Lambda>1$, this condition can be met by
increasing $\Lambda$ (instead of $N$), since $\tau_{\st}$
increases with $\Lambda$, as discussed in Sec.\ \ref{sec:curr}.
Constant currents corresponding to perfect conductance can, in
principle, be reached for $\Lambda$ values large enough so that
$\tau_{\st}(\Lambda) > \tau_K$. In this sense, this requirement
marks the regime for which the steady state of this problem can be
numerically simulated.

Once the condition $\tau_{\st}(\Lambda) > \tau_K$ is fulfilled,
one still needs to run the tDMRG algorithm over time scales of
order $\tau_K$ to obtain the Kondo plateau. Thus, for higher
values of $U/\Gamma$ ({\it i.e.}, higher $\tau_K$), calculations
over longer time scales are necessary in order to reach nearly
perfect conductance plateaus in $J(\tau)$.\cite{wingreen93} This
is, due to the entanglement growth in a global quench,
\cite{dechiara06,calabrese07} the true limitation of the method
for Kondo problems. Fortunately, the entanglement growth turns out
to be softer at large values of $U/\Gamma$ (see
App.~\ref{app:comp}), enabling us to observe $G\approx G_0$ for up
to $U/\Gamma\sim 6$ and relatively short ($N=32$) chains.

This argument is further supported by quantitative estimates for
$\tau_K$ obtained with NRG (inset in Fig.\ \ref{fig:G_tp_1dot}).
We performed NRG calculations for the Anderson model and extracted
the Kondo temperature (and thus $\tau_K$) from the  magnetic
susceptibility curves for different values of $U/\Gamma$.
\cite{krishnamurthy80a} For the parameters in Fig.\
\ref{fig:G_tp_1dot},  we obtain $\tau_K<\tau_{\st}$ in the regime
where nearly perfect conductance is seen in the tDMRG curves
($\tau_K \simeq 16$ for $U/\Gamma=3.125$ and $\tau_K \simeq 55$
for $U/\Gamma=5$, in units of $\hbar/t_0$). \cite{NRGdetails} For
higher values of $U/\Gamma$, $\tau_K$ becomes exponentially large.
In particular, for $U/\Gamma \sim 7$, $\tau_K$ calculated from NRG
becomes of the order of the maximum time scales used in our tDMRG
simulations. This explains the noticeable deviation of $G$ from
the Kondo value for $U/\Gamma > 7$.

In short, the tDMRG results obtained with $\Lambda>1$ constitute a
considerable improvement over the $\Lambda=1$ case, as shown in
Fig. \ref{fig:G_tp_1dot} for $N=32$.  This plot illustrates the
range of parameters for which the Kondo regime is accessible with
tDMRG and $\Lambda>1$, as well as the typical system sizes.

\section{Finite bias}
\label{sec:bias}

In this Section, we address  the case of  a current through a  quantum
dot in  the   Kondo regime  driven    by a finite bias.   Although   a
comprehensive theoretical understanding  of this nonequilibrium regime
is yet  to be achieved, one  commonly expects the applied bias $\Delta
V$  to disrupt the   Kondo state for bias  voltages  larger than $T_K$
while Kondo-like properties are only marginally affected for $\Delta V
\ll T_K$.\cite{fujii03,jakobs07,kirino08,weiss08}    We  investigate the
transport   properties of  the  system in   these two  regimes. In the
following, we fix the parameters  to $U/\Gamma=3.125$ at the particle-hole
symmetric   point  ($V_g=-U/2)$,   for  which  we   have independently
determined $T_K$ from NRG calculations\cite{NRGdetails} (see the inset
in Fig. \ref{fig:G_tp_1dot}).

Qualitatively, we   expect the  linear   regime (i.e.,  nearly perfect
conductance) to extend up to biases $\Delta V\sim T_K$.  The current
versus bias curve should then smoothly  drop to zero at larger biases.
We argue that, in the  spirit of the previous  sections, in the regime
$\Delta V < T_K$, the finite-bias regime should  best be explored with
discretized leads, for which the   Kondo   state is better described.

By contrast, in the opposite   limit of $\Delta V\gg T_K$,
the scan in bias will need the high-energy features  (such as the band
curvature in the leads) to be well resolved as the contribution to the
current from states with energies within the  Fermi levels in the left
and the right leads   becomes  important. Therefore, at  $\Delta  V\gg
T_K$,   the  best   approach  is to  use   $\Lambda=1$  in  the  tDMRG
calculations and  subsequently perform  a finite-size scaling analysis
of the average currents.

\begin{figure}[tbp]
\includegraphics[clip,width=1\columnwidth]{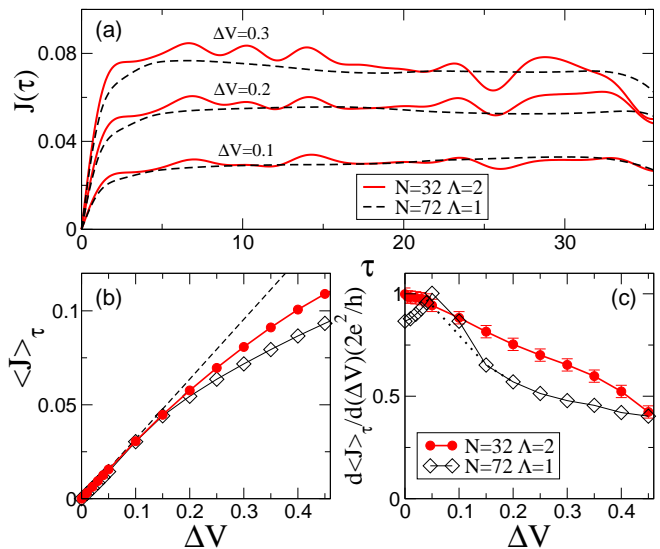}
\caption{ (color online) (a) Current $J(\tau)$ versus time 
for $\Delta V=  0.1,0.2,0.3$ and $\Lambda=1$ ($N=72$) and  $\Lambda=2$
($N=32$). (b)  Average current  $\langle  J \rangle_{\tau}$  and  conductance
$d\langle  J \rangle_{\tau}/d(\Delta V)$ as a  function  of bias.  The dashed
line is   $\langle J\rangle_{\tau}   =  G_0  \Delta  V$.   (c)   Differential
conductance for  $\Lambda=1$  ($N=72$) and $\Lambda=2$  ($N=32$).  
The dotted line represents an interpolation between the low bias regime
(better described  by $\Lambda=2$) and   the   high bias  one   (better
described by $\Lambda=1$).
}
\label{fig:finite-bias}
\end{figure}

Our   results  are      illustrated   in   Fig.~\ref{fig:finite-bias}.
Fig.~\ref{fig:finite-bias}(a)   shows  the  current versus  time for
different  bias   values  and   $\Lambda=1$, $N=72$ and   $\Lambda=2$,
$N=32$.  As   a  general  feature,  the  average   current  $\langle J
\rangle_{\tau}$ increases with the bias $\Delta V$ in all cases, also
seen in Fig.   \ref{fig:finite-bias}(b). Moreover, it is evident  that
for      the      values     of    $\Delta     V$     depicted      in
Fig.~\ref{fig:finite-bias}(a),  runs   with  either  $\Lambda=1$    or
$\Lambda=2$ give a {\it qualitatively} similar behavior.

A  more quantitative analysis of  the $\langle J \rangle_{\tau}(\Delta
V)$ curves   obtained from   the tDMRG  data   is  presented in  Figs.
\ref{fig:finite-bias}(b) and \ref{fig:finite-bias}(c).  Deviations from perfect conductance
($\langle J\rangle_{\tau}  = G_0 \Delta  V$)  are clear at large  bias
(Fig.  \ref{fig:finite-bias}(b)).  At   small biases, the  results are
better  visualized by    numerically  calculating   the   differential
conductance   $d\langle  J   \rangle_{\tau}/d(\Delta  V)$,   shown  in
Fig.~\ref{fig:finite-bias}(c). For  small $\Delta V$, this quantity is
equivalent to the linear conductance.

For $\Lambda=1$ and  $N=72$ sites,  we see  that the corresponding  $d
\langle J \rangle_{\tau}/d   (\Delta V)$ curve  is substantially below
$G_0$  for  $\Delta  V   \lesssim T_K$   in sharp  contrast   with the
$\Lambda=2$ results, emphasizing the importance of using Wilson chains
at small  biases.  At large  biases,  we have  performed a finite-size
scaling analysis   with $N\leq 80$,   and  we conclude  that  the data
displayed  in  Fig.~\ref{fig:finite-bias}(c)  are converged   down  to
$\Delta V \gtrsim 0.15$ with an uncertainty of $\delta \left[d \langle
J \rangle_{\tau}/d (\Delta V)\right] \sim 0.01$.

In an  intermediate bias regime  of  $T_K  \lesssim \Delta V  \lesssim
0.15$, the $\Lambda=1$  results    are still plagued    by finite-size
effects,  while  the   $\Lambda=2$  results overestimate  the   actual
steady-state  currents.  In this regime,  a discretization scheme that
interpolates between linear and   a logarithmic  discretization  will
likely  be the   best  choice.  Qualitatively,  one  can estimate  the
expected result by a linear  interpolation between the $\Lambda=1$ and
$\Lambda>1$  data,    as  illustrated   by      the  dotted  line   in
Fig.~\ref{fig:finite-bias}(c)

Finally,    note  that the   finite-bias   regime of a single-impurity
Anderson  model  has been    studied  with tDMRG  by Kirino   {\it  et
al.}~(Ref.~\onlinecite{kirino08}) for  $U/\Gamma \lesssim   2.8$   and
system sizes of up to  $N=64$ (all at  $\Lambda=1)$. Here, we slightly
exceed  that   regime by working    at  $U/\Gamma=3.125$, and  we also
consider larger system sizes of  $N=64,72,80$.  Our key point, though,
is  that tDMRG runs at $\Delta  V\lesssim T_K$ notoriously improve and
correctly capture  Kondo   correlations when   performed with   Wilson
chains
.

\section{Summary}
\label{sec:sum}
In this paper, we  applied the tDMRG method  to the study of transport
through a  quantum dot   coupled   to  noninteracting  leads  with   a
logarithmic  discretization.  This yields  a considerable  improvement
over tDMRG studies with real-space  tight-binding leads, as it extends
the  parameter space in which known  exact  results can be reproduced.
One of the main advantages of the approach  is that smaller chains are
sufficient to obtain the expected result  of a perfect conductance for
the single-impurity problem at particle-hole symmetry.

In  spite of the challenges imposed  by the  longer time scales needed
for   the  description of  the Kondo  regime,  the  study of transport
properties in  nanostructures with time-dependent DMRG  brings several
advantages over other methods: it is straightforward to adapt codes to
more complicated geometries and time-dependent Hamiltonians, including
correlation effects    in the leads,   as  well  as systems   far from
equilibrium,  such   as transport  beyond   the linear response regime
(finite-bias). \cite{kirino08} For the latter  example, we presented a
case study at an intermediate $U/\Gamma$ to argue that for bias values
($\Delta   V   \lesssim T_K$),   a   logarithmic discretization should
preferentially be  used, while at    large  bias, a regime  in   which
high-energy  features of the  leads   dominate,  a  tDMRG study   with
$\Lambda=1$ and a finite-size scaling analysis yields better results.

Furthermore, we believe the general concept of utilizing a logarithmic
discretization in tDMRG  can have a  broader range of applicability in
the description of Kondo systems.   In particular, it  can play a  key
role  in  the    calculation of  nonequilibrium    dynamic correlation
functions,      as      recently     highlighted      in     NRG-based
approaches. \cite{anders08}

\acknowledgments

We sincerely thank K.~A.    Al-Hassanieh, I.~P.  McCulloch,  G.  Roux,
H. Onishi, and U.  Schollw\"ock for  fruitful discussions and valuable
comments on  the  manuscript.  Research at  ORNL is   sponsored by the
Division of Materials Sciences and Engineering, Office of Basic Energy
Sciences, U.S.  Department of Energy, under contract DE-AC05-00OR22725
with    Oak Ridge  National   Laboratory,   managed  and operated   by
UT-Battelle, LLC.  E.D.,  and L.D.d.S.  are supported  in  part by NSF
grant DMR-0706020. F.H.-M.   acknowledges support from the DFG through
FOR 912. G.B.M. and C.A.B. acknowledge support from NSF (DMR-0710529).
E.V.A.   acknowledges  support  from  Brazilian   agencies FAPERJ  and
CNPq (CIAM 490865/2006-2)

\appendix

\section{Noninteracting case}
\label{app:u0}

\begin{figure}[tbp]
\includegraphics[clip,width=0.9\columnwidth]{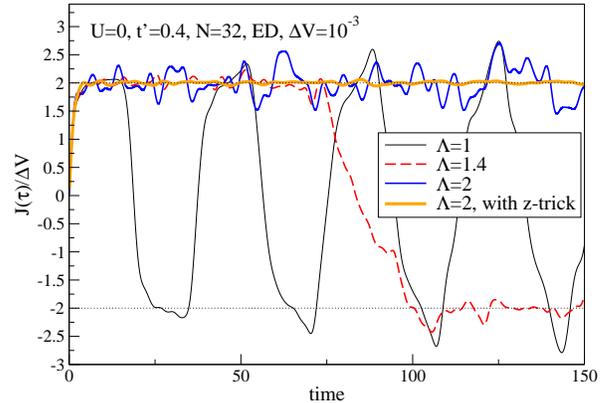}
\caption{ (color online) Current $J(\tau)$ as a function of time
for the noninteracting case ($U=0$, $V_g=0$, $t'=0.4$, $N=32$) for
different values of $\Lambda=1,1.4,2$, calculated with exact
diagonalization. }
\label{fig:u0}
\end{figure}

In this appendix, we present exact diagonalization results for the
time-dependence of the current Eq.~(\ref{eq:j}) for a chain of
$N=32$ sites and $t'=0.4$. Figure~\ref{fig:u0} shows $J(\tau)$ vs.
time for $\Lambda=1,1.4,2$. Some main features induced by the
discretization discussed in Sec.~\ref{sec:curr} are already
present in the noninteracting case: (i) as $\Lambda$ is increased,
the sign reversal of the current occurs at much later times and
(ii) while the average current in the $U=0$ case remains at
$G=G_0$, the dispersion around this mean value is significantly
enhanced by a $\Lambda>1$.

More specifically, by taking averages over suitable time
intervals, we find $G/G_0=(1.00\pm 0.04)$  and $G/G_0=(1.00\pm
0.13)$, for  $\Lambda=1.4$ and 2, respectivley. The deviations are
computed from Eq.~(\ref{eq:deltaG}). We thus conclude that the
oscillations seen in Fig.~\ref{fig:G_NLambda}
 in the interacting case are due to the discretization.

Moreover,  from  the  noninteracting  case  we learn  that  using  the
logarithmic discretization, one manages to reproduce the average quite
well. Once one has achieved that in the interacting case for a pair of
$(N,\Lambda)$,  we expect  that the  oscillations will  be  reduced by
increasing  $N$  and  decreasing  $\Lambda$  at the  same  time  in  a
controlled  way  such  that  the  average  current  $J(\tau)$  remains
constant.

An  alternative way  of suppressing  the oscillations  induced  by
the discretization       is      to       exploit       the
so-called z-trick.\cite{oliveira94,bulla08,anders06}  Indeed, this
works quite well: the thick solid line  in Fig.~\ref{fig:u0} is
obtained by using the  z-trick for $\Lambda=2$,  which clearly
improves the  data quality over  the  simple $\Lambda=2$  curve.
However,  it  turns out  to  be necessary  to average  over  many
values of  z:
results shown in Fig.~\ref{fig:u0} were obtained by averaging over
forty $J(\tau)$-curves with z-values ranging from 0 to 1. In
practice, this makes this procedure numerically expensive for
tDMRG calculations.

\section{Computational aspects}
\label{app:comp}

\begin{figure}[tbp]
\includegraphics[clip,width=0.9\columnwidth]{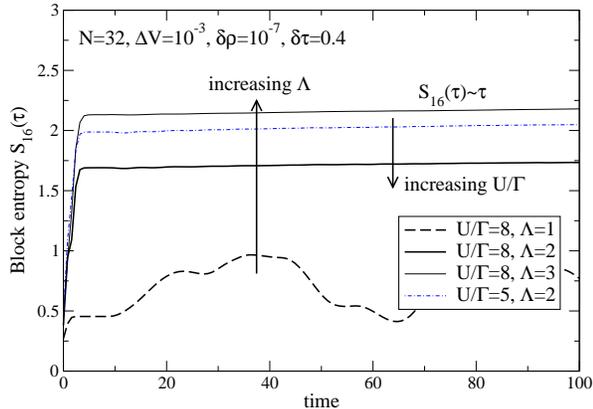}
\caption{ (color online)  Block entropy $S_l$ for a block of length $l=16$ vs. time
for $U/\Gamma=5$ ($\Lambda=2$) and $U/\Gamma=8$ ($\Lambda=1,2,3$).
$N=32$ and $\Delta V=10^{-3}$, the time steps $\delta \tau$, and the truncated weight $\delta \rho$ are given in the legend.
}
\label{fig:entropy}
\end{figure}

As a key result of this work, we have argued that using the logarithmic discretization
much smaller chains than in the $\Lambda=1$ case can be used to obtain
equally good, if not better, results for the conductance. This suggests a gain in
the computational costs needed to obtain the numerical results, by  reducing the
required system sizes roughly by a factor 4.

For a more stringent estimate of the computational efficiency, we
consider the entanglement growth during the
time-evolution.
We measure
this quantity
by computing  the \textit{block entropy} $S_l$, associated with
the  reduced  density matrix  $\rho_{l}$  of  a DMRG block  of
length $l$:
\begin{equation}
 S_l=- \langle \rho_{l} \mathrm{ln} \rho_{l}\rangle  \,.
\label{eq:entropy}
\end{equation}

The reduced density matrix
$\rho_{l}$ is obtained at each step by  dividing the DMRG  chain
(the so-called super-block) into ``system'' (size $l$) and
``environment'' parts,  and tracing  out the environment's degrees
of  freedom (see, {\it  e.g.}, Ref.~\onlinecite{schollwoeck05} for
a discussion of the DMRG method). An increase in $S_l$ renders  a
simulation inefficient as time or system size grows,  since more
DMRG states need  to be kept in  order to keep the truncation
error below  a given threshold.

In Fig.~\ref{fig:entropy},  we plot the  block entropy for a  block of
length $l=16$  as a  function of time,  for two  different values of $U/\Gamma=5$
($\Lambda=1$) and 8 ($\Lambda=1,2,3$).  Typically, the entropy rapidly
increases at short times, but at times $\tau \gtrsim 3$, it exhibits a
linear increase in time for $\Lambda>1$, $S_l  \propto \tau$. This is the expected behavior for a global
quench,\cite{dechiara06,calabrese07,schoenhammer08} yet it is a non-obvious one here  as
the excitations in the leads do not travel at a constant velocity (see Fig.~\ref{fig:ChargeProfile_1dot}(b)).
The oscillations in $S_l(\tau)$ seen in the $\Lambda=1$ case (dotted line)
are due to the sign reversal of the current.
The key  point is
that this linear increase is slow, {\it i.e.}, the prefactor is small.
During the  time interval $\tau \in \lbrack  5,100\rbrack$, $S_l$ only
grows by  a few per  cent.  This is  ultimately the reason why  we can
push our tDMRG runs to times long enough to reach the steady state for
$U/\Gamma \lesssim 6$ at moderate numerical costs, especially
since the entanglement growth is the weaker the larger $U/\Gamma$ is.

We thus observe that the entanglement growth, {\it i.e.}, the increase
in the entropy $S_l$, depends on both $\Lambda$ and $U/\Gamma$.
Yet, it  is fortunate
that in the case where longer times are needed in order to capture the
steady-state current (large $U/\Gamma$) the increase in $S_l$ is weaker.

It is illustrative to give  an example on what the entanglement growth
implies in practice  for the numerical effort when  working at a fixed
truncation  error $\delta  \rho$.   For $U/\Gamma=3.125$,  we find  it
sufficient  to  keep $m\sim280$  states  at  $\Lambda=1$  in order  to
ensure a maximum   truncated weight of $\delta\rho\sim 10^{-7}$  on a chain  of $N=32$ sites,
compared  to  $m\sim1000$  at  $\Lambda=\sqrt{2}$ and  $m\sim1600$  at
$\Lambda=3$  (both numbers  refer  to times  $\tau  \lesssim 30$ with  $\delta
\tau=0.05$).   At a  larger $U/\Gamma$,  say $12.5$,  this  relaxes to
$m\sim200$ and $m\sim400$, for $\Lambda=1$ and $\Lambda=\sqrt{2}$, 
respectively.

We  finally  comment on  the  generic  tDMRG  errors, the  accumulated
truncation error, and the Trotter error. These are not independent: the
smaller   the   time  step,   the   faster   truncation  errors   will
accumulate.\cite{gobert05}  We  justify our  choice  of parameters  by
considering the numerically worst  case, {\it i.e.}, a small $U/\Gamma
\sim 3$.

For most calculations, keeping a maximum of $m=660$ states up to
times of order $\tau\sim 50$ is sufficient to keep the truncation
error below $10^{-7}$ ($\sim 3 \times 10^{-7}$ in a few sweeps).
More importantly, we have checked that, keeping up to $m=1600$ states,
the current $J(\tau)$ is practically converged: for the case of Fig.~\ref{fig:G_NLambda}(c) 
and $\Lambda=2$, the maximum relative change in $J(\tau)$ is  $\sim 1$\%,
comparing runs with $\delta \rho =10^{-7}$ and $\delta \rho= 10^{-8}$. In
addition, we have calculated the forth-back error\cite{gobert05}
to validate that this is a sufficiently small discarded weight for
our purposes, in which the oscillations cause the dominant fluctuation 
around the current's average (see App.~\ref{app:u0}).

We have further  checked our tDMRG with the  chosen parameters against
exact diagonalization  for the noninteracting case to  make sure that
the so-called run-away time\cite{gobert05} is not the limiting factor in our case.



\begin{thebibliography}{10}
\bibitem{NRGdetails}
$T_K$ is calculated within NRG in units of the half-bandwidth $D$ of the
leads. A comparison of the parameters entering NRG and DMRG
calculations can be made by identifying $D=2t_0$ where $t_0$ is the
tight-binding hopping in the DMRG chain in the continuum limit and
$\Lambda=1$.
\end{thebibliography}
\end{document}